\documentclass[epj]{svjour}
\usepackage{graphics,epsfig}
\begin{document}
\renewcommand{\thefootnote}{\alph{footnote}}
\title{S-factor of $^{14}\mathrm{N}(\mathrm{p},\gamma)^{15}\mathrm{O}$ at astrophysical energies
\thanks{LUNA collaboration, Supported in part by INFN, GSI (Bo-Rol 125/2) and BMBF (05CL1PC1-1)}
\author{ G.Imbriani     \inst{1},
         H.Costantini   \inst{2},
         A.Formicola    \inst{3}, 
         A.Vomiero      \inst{4}, 
         C.Angulo       \inst{5}, 
         D.Bemmerer     \inst{6}, 
         R.Bonetti      \inst{7}, 
         C.Broggini     \inst{6},
         F.Confortola   \inst{2}, 
         P.Corvisiero   \inst{2}, 
         J.Cruz         \inst{8}, 
         P.Descouvemont \inst{9}, 
         Z.F\"ul\"op      \inst{10},
         G.Gervino      \inst{11},
         A.Guglielmetti \inst{7}, 
         C.Gustavino    \inst{3}, 
         Gy.Gy\"urky       \inst{10},
         A.P.Jesus      \inst{8}, 
         M.Junker       \inst{3},
         J.N.Klug       \inst{12},
         A.Lemut        \inst{2}, 
         R.Menegazzo    \inst{6}, 
         P.Prati        \inst{4}, 
         V.Roca         \inst{1}, 
         C.Rolfs        \inst{12},
         M.Romano       \inst{1}, 
         C.Rossi-Alvarez\inst{6}, 
         F.Sch\"umann   \inst{12},
         D.Sch\"urmann  \inst{12},
         E.Somorjai     \inst{10},
         O.Straniero    \inst{13},
         F.Strieder     \inst{12},
         F.Terrasi      \inst{14},
         H.P.Trautvetter\inst{12}  } 
%
}                     
\offprints{G.Imbriani, gianluca.imbriani@na.infn.it, 
Universit\'a di Napoli "Federico II", Dipartimento di Fisica, Complesso Universitario di Monte Sant'Angelo, 
Via Cintia ed. G, 80126 Naples, Italy - tel:+39081676853}          
\institute{Universit\'a  di Napoli "Federico II", Dipartimento di Fisica and INFN, Napoli, Italy                 
  \and Universit\'a  di Genova, Dipartimento di Fisica and INFN, Genova, Italy                                
  \and Laboratori Nazionali del Gran Sasso dell INFN, Assergi, Italy                                          
  \and Universit\'a  di Padova, Dipartimento di Fisica and INFN Legnaro, Italy   
  \and Centre de Recherches du Cyclotron, Universit\'e catholique de Louvain, Louvain-la-Neuve, Belgium
  \and INFN, Padova, Italy                                                                                    
  \and Universit\'a  di Milano, Dipartimento Di Fisica and INFN, Milano, Italy                                        
  \and Centro de Fisica Nuclear da Universidade de Lisboa, Lisboa, Portugal                                
  \and Physique Nucl\'eaire Th\'eorique et Physique Math\'ematique, CP 229, Universit\'e Libre de Bruxelles, Brussels, Belgium
  \and Atomki, Debrecen, Hungary                                                                             
  \and Universit\'a di Torino, Dipartimento di Fisica Sperimentale and INFN, Torino, Italy                   
  \and Ruhr-Universit\"at Bochum, Institut f\"ur Physik mit Ionenstrahlen, Bochum, Germany                           
  \and Osservatorio Astronomico Collurania, Teramo and INFN Napoli, Italy                                    
  \and Seconda Universit\'a  di Napoli, Dipartimento di Scienze Ambientali, Caserta and INFN, Napoli, Italy  
}
\date{Received: date / Revised version: date}
%
\abstract{
The astrophysical S(E) factor of $^{14}\mathrm{N}(\mathrm{p},\gamma)^{15}\mathrm{O}$ has been measured 
for effective center-of-mass energies between E$_{eff}$= 119 and 367 keV at the LUNA facility using 
TiN solid targets 
and Ge detectors. The data are in good agreement with previous and recent work at overlapping energies. 
R-matrix analysis  reveals that due
to the complex level structure of $^{15}\mathrm{O}$ the extrapolated S(0) value is model 
dependent and calls for additional experimental efforts to reduce the present uncertainty 
in S(0) to a level of a few percent as required by astrophysical calculations.
} 
\titlerunning{S-factor of $^{14}\mathrm{N}(\mathrm{p},\gamma)^{15}\mathrm{O}$ at astrophysical energies}
\authorrunning{G. Imbriani, et al}
\maketitle
\section{Introduction}
\label{intro}
\label{Intro}

\begin{figure*}
  \includegraphics[width=15cm]{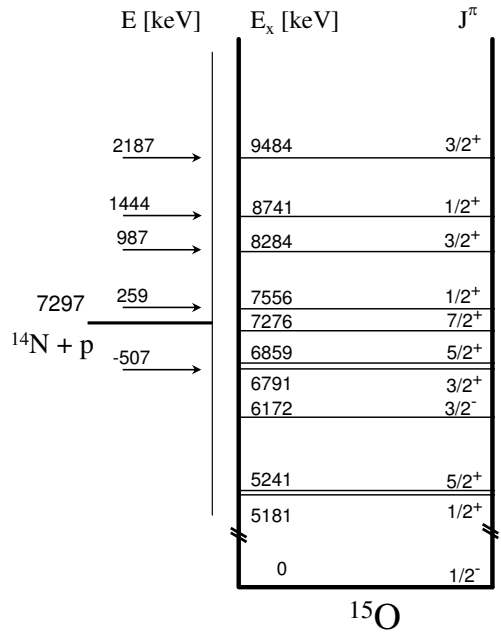}
\caption{Relevant level scheme of $^{15}O$ near the $^{14}\mathrm{N}(\mathrm{p},\gamma)^{15}\mathrm{O}$ threshold.}
\label{structure}       
\end{figure*}

The capture reaction $^{14}\mathrm{N}(\mathrm{p},\gamma)^{15}\mathrm{O}$, the slowest process in the hydrogen burning 
CNO cycle \cite{rol88}, is of high astrophysical interest as its reaction rate influences sensitively the
age determination of globular clusters \cite{imbriani04} and the solar neutrino spectrum \cite{bp04,d04}. 
The capture cross section needs to be known down to $E_0 = 30$ keV (the Gamow peak in core H-burning stars), 
which is far below the low-energy limit of direct $\gamma$-ray measurements, i.e. the center-of-mass energy 
$E = 240$ keV{\footnote {In this work all proton energies are taken in the center-of-mass system,
except where quoted differently.}} \cite{schroeder87}. 
Thus, the data had to be extrapolated over a large energy gap leading to a substantial uncertainty for the 
astrophysical S-factor at zero energy, S(0). 
According to the data and analyses of Schr\"oder et al \cite{schroeder87}, there are two major and 
nearly equal contributions to S(0): the direct capture (DC) to the 6.79 MeV state in $^{15}O$ and the capture to the 
ground state (gs) in $^{15}O$. 
The latter process is enhanced due to a subthreshold resonance at $E_R=-507$ keV (Fig. \ref{structure}), 
the width of which was taken as a free parameter in the fit \cite{schroeder87}. 
The extrapolation led to $S_{tot}(0)=3.20\pm0.54$ keVb, with $S_{gs}(0)=1.55$ keVb and $S_{6.79}(0) = 1.41$ keVb.

Subsequently, the data of Schr\"oder et al \cite{schroeder87} were analyzed by Angulo and Descouvemont \cite{angulo01} using an 
R-matrix approach \cite{lane58}. 
Contrary to the extrapolation by Schr\"oder et al \cite{schroeder87} for capture to the ground state, they reported a negligible 
contribution $S_{gs}(0)=0.08$ keVb due to a smaller total width of the subthreshold resonance and thus they suggested 
$S_{tot}(0)=1.77\pm0.20$ keVb. A smaller width of the 6.79 MeV state was supported by a lifetime measurement via the 
Doppler-shift method \cite{bertone01} and by a Coulomb excitation measurement \cite{yamada04}. 
     
Both $S_{gs}(0)$ and $S_{6.79}(0)$ are dominated by E1 capture mechanisms. Based on a measurement of the analysing power at 
$E_p = 270$ keV, Nelson et al. \cite{TUNL03} reported a small contribution (about 4\%) of an M1 capture to the overall capture process, 
but this single data point did not significantly improve the situation. 
The available data clearly demonstrate the large uncertainty affecting the extrapolations over a 
large energy gap in a nucleus, 
such as $^{15}O$, with a complex level structure, a problem calling for new direct $\gamma$-ray measurements towards lower energies.

The LUNA (\rm{L}aboratory \rm{U}nderground for \rm{N}uclear \rm{A}strophysics) collaboration started in 2001 a reinvestigation of 
$^{14}\mathrm{N}(\mathrm{p},\gamma)^{15}\mathrm{O}$ (using TiN solid targets and Ge detectors) with a particular emphasis on $S_{gs}(E)$. 
The data obtained down to $E_{p} = 140$ keV \cite{formicola04} confirmed the small contribution of the ground state capture 
giving $S_{tot}(0)=1.7\pm0.1\pm0.2$ keVb. The R-matrix analysis of these low-energy data included the high-energy 
data of Schr\"oder et al. \cite{schroeder87}  and the results of the ANC (Asymptotic Normalization Coefficient) 
method \cite{bertone02,mukhamedzhanov03}.
At the completion of the LUNA work, another set of low-energy data became available at 
E$_p$ = 155 to 524 keV \cite{ru05}, which is in excellent agreement 
with the LUNA data in the overlapping energy range. 
Here, we report on the LUNA solid target work, while additional details can be found in \cite{forthes,heide03,klug05}.

\section{Experimental apparatus}
\label{apparatus}

\subsection{Accelerator}
\label{accelerator}

\begin{figure*}
  \includegraphics[width=15cm]{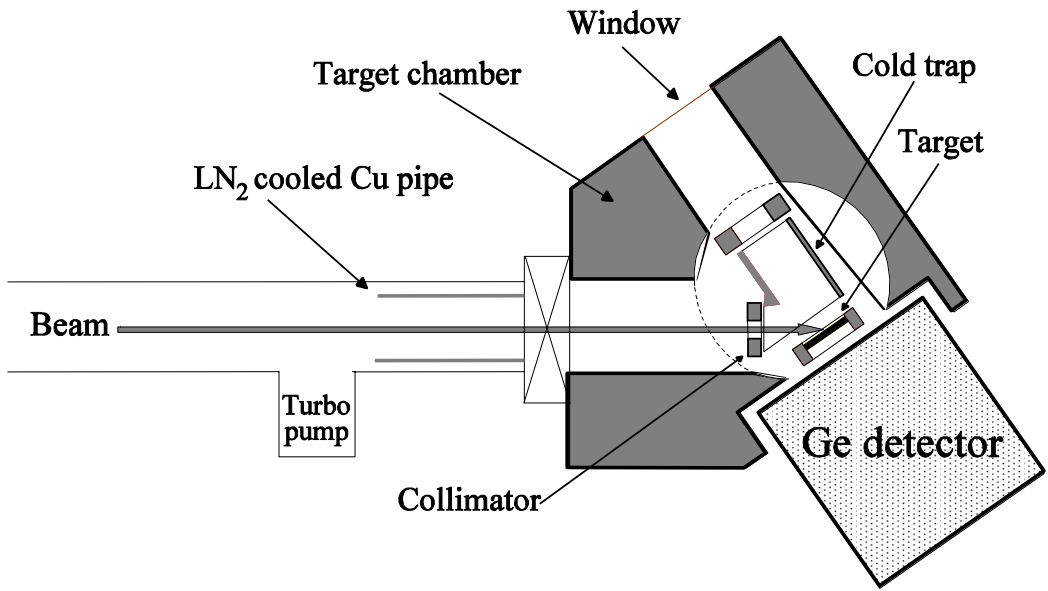}
\caption{Schematic diagram of the setup used in geometry 1.}
\label{geom1}       
\end{figure*}

The 400 kV LUNA accelerator \cite{formicola03} provided a proton beam on target of up to $500$ $\mu$A. 
The absolute beam energy was known with an accuracy of $300$ eV, and the energy spread 
and the long-term energy stability were observed to be $100$ eV and $5$ eV/h, respectively. 
The beam current on target was monitored by a current integrator with an estimated uncertainty of $2~\%$. 
The targets were directly water-cooled. In the target chamber (Fig. \ref{geom1}) the target ladder was 
surrounded by an 
electrically insulated collimator biased to -300 V to suppress secondary electrons. 
A uniform rectangular beam spot (4 by 4 cm) was produced within the target area by magnetic wobbling of the beam. 
In order to prevent build-up of impurities on the target, a LN$_2$-cooled Cu cold finger was used. 

\subsection{Detectors and electronics}
\label{detect}

The laboratory's 1400 m rock cover reduces the background rate in the Ge detectors by more than three orders of magnitude 
at $E_\gamma > 5$ MeV \cite{bemmerer05} compared with a detector placed at the surface of the earth. 
It is in this $E_\gamma$ region that the capture $\gamma$-ray lines of secondary transitions (5.18, 6.17, and 6.79 MeV) and the 
ground state transition (about 7.5 MeV) are located (Fig. \ref{structure}). In the experiments two different setups were used.
\begin{list}{}
\item Geometry 1 (Fig. \ref{geom1}): The target was oriented with its normal at 55$^\circ$ with respect to the beam direction. 
The capture  $\gamma$-rays were observed with a HPGe-detector (126$\%$ relative efficiency), 
whose front face and target were parallel with a distance $d=1.5$ cm. The detector was surrounded by 5 cm of lead, which 
reduced the background in the low-energy range of the primary transitions by a factor of 10. The setup was used for the 
measurement of excitation functions, while for the determination of the detector efficiency and of summing effects
the distance between target and detector was increased in discrete steps up to $d=20.5$ cm.
\item Geometry 2: The target normal was oriented at $19^\circ$ with respect to the beam direction. Using a different target 
chamber three HPGe-detectors were positioned around the target: the $126\%$ detector was placed at $0^\circ$, 
while the other two detectors ($120\%$ and $108\%$ relative efficiencies) were placed at $90^\circ$ and $125^\circ$. 
The distance between the detector end cap and the target was 7 cm for all detectors. 
The setup was used in the measurement of Doppler shifts, excitation energies, and angular distributions.
\end{list}

\subsection{Targets}
\label{targets}
    
\begin{figure*}
  \includegraphics[width=15cm]{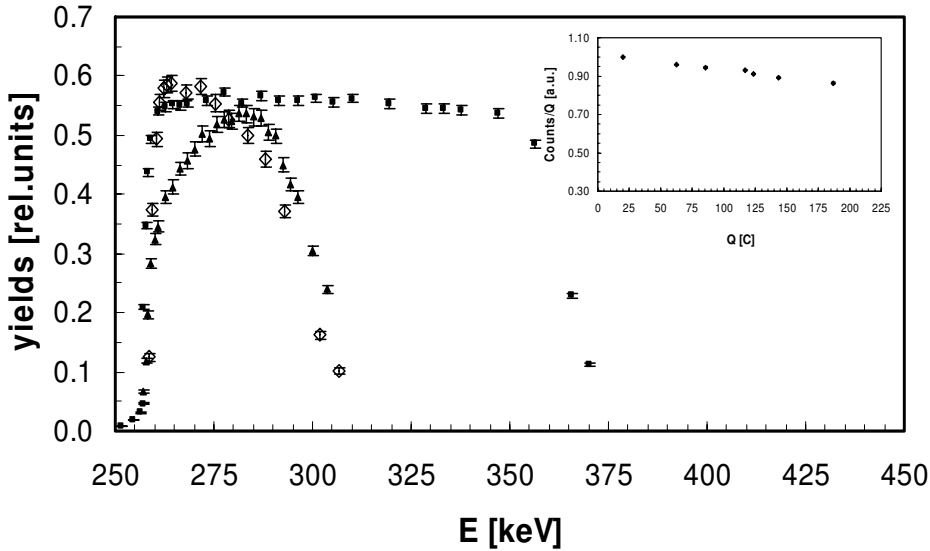}
\caption{Excitation function of $^{14}\mathrm{N}(\mathrm{p},\gamma)^{15}\mathrm{O}$ near the $E_R = 259$ keV resonance 
using different targets: squares = sputtered target, triangles = implanted target, diamonds = evaporated target. 
The insert shows the thick-target yield as a function of accumulated charge Q on the sputtered target.}
\label{target}       
\end{figure*}

The $^{14}N$ solid targets had to fulfill several requirements: (i) uniform depth distribution starting at the surface, 
(ii) thickness large enough that possible sputtering effects at low energies and over long running times were 
negligible for the yield measurements, (iii) stability against high beam power, and (iv) low concentration of 
impurities of light elements. The targets were produced either by implantation of a $^{14}N$ ion beam into a suitable 
backing, or by evaporation of a Ti layer on a specific backing and heating it in a N$_{2}$ atmosphere, or by sputtering of TiN on a backing. 
The resulting quality of the targets was investigated using the $E_R = 259$ keV resonance. 
Fig. \ref{target} shows the results out of a large number of tests \cite{forthes} performed with different backings 
and target thicknesses. The evaporated and sputtered targets showed a steep rise in the excitation 
function at the resonance energy, with a slope comparable to the convolution of beam resolution 
and resonance width. In the implanted target the surface region was not saturated. The step heights to the plateau 
were similar for all targets indicating a similar stoichiometry at saturation. 
Our conclusion was that the sputtered targets of TiN (with a typical thickness of 80 keV, Ta-backing) had the most 
uniform number density profile and could withstand many days of intense beam bombardment without a significant deterioration. 
Typically, a new TiN target was installed after a running time of 1 week, which corresponded to an accumulated charge of about 200 C. 
During this time we observed a small decrease of the plateau-yields due to the proton implantation (insert of Fig. \ref{target}). 
The stoichiometry of the TiN layer - used for the determination of the strength of the $E_R = 259$ keV resonance - was 
measured via Rutherford Backscattering Spectrometry using a 2.0 MeV $^4$He beam in the INFN National Laboratory of Legnaro, resulting in $Ti/N = 1/(1.08 \pm 0.05)$. 
To measure the capture cross section above the $E_R = 259$ keV resonance we used evaporated targets, since 
we needed here thinner targets to minimize the influence of the resonance: their absolute stoichiometry did not 
need to be precisely known because of the normalization procedure adopted (section \ref{primary}).

\subsection{Beam induced background}
\label{background}

\begin{figure*}
  \includegraphics[width=15cm]{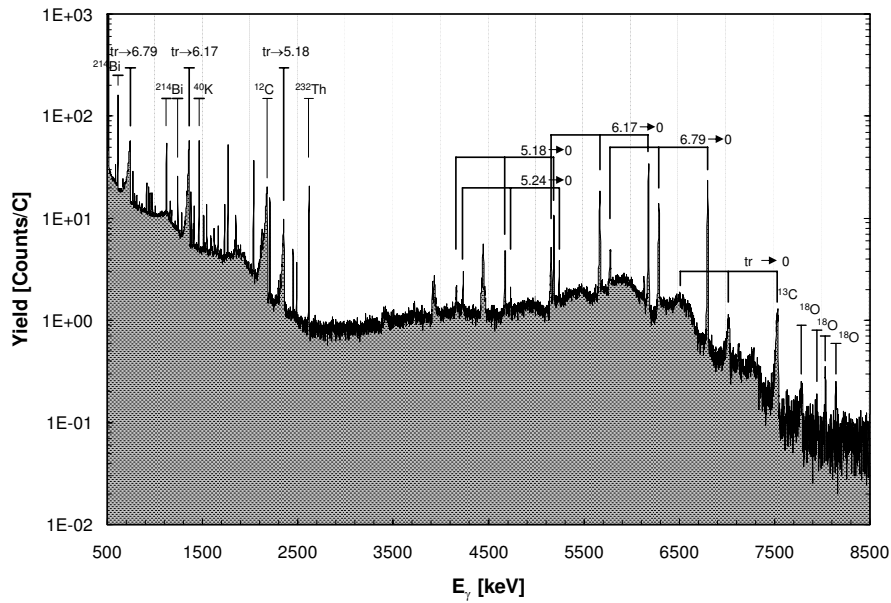}
\caption{Spectrum obtained at $E_p = 250$ keV in geometry 1 over an accumulated charge of $87$ C. 
The background above $E_\gamma  = 8$ MeV is mainly due to a small $^{11}$B contamination.}
\label{230}       
\end{figure*}

\begin{figure*}
  \includegraphics[width=15cm]{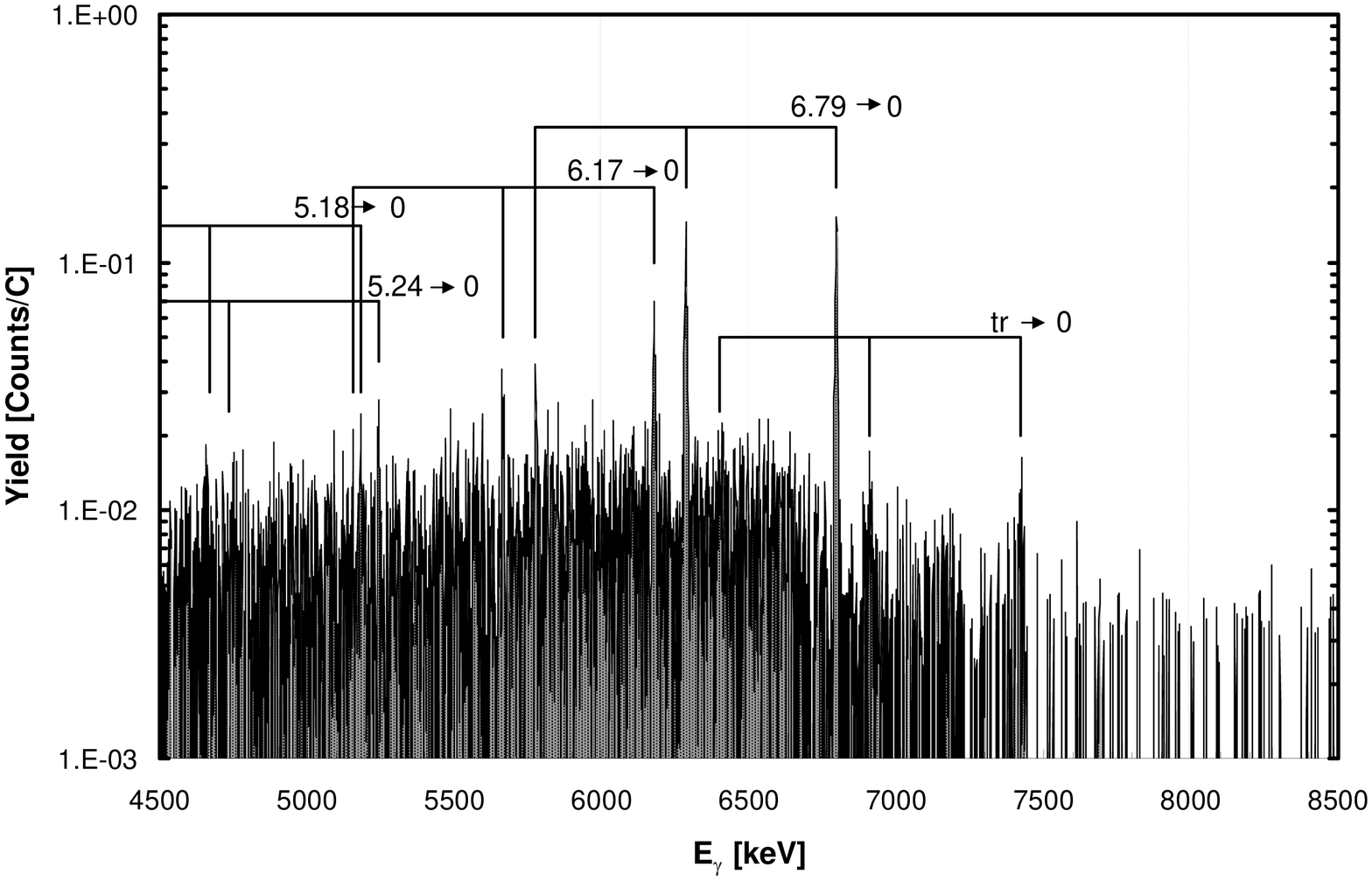}
\caption{Spectrum obtained at E$_p$ = 140 keV in geometry 1 over an accumulated charge of $210$ C.}
\label{130}       
\end{figure*}

The presence of light-element impurities can produce intense $\gamma$-ray background lines due to their 
relatively low Coulomb barrier and/or high cross sections. In particular  $\gamma$-lines from reactions on 
$^{11}B$, $^{12}C$, $^{13}C$, $^{18}O$ and $^{19}F$ have been investigated and their intensity was minimized 
by a proper choice of target backing and preparation procedures \cite{strieder03}, where the sputtered target was best. 
A sample  $\gamma$-ray spectrum obtained 
at $E_p = 250$ keV is shown in Fig. \ref{230}: the primary capture transitions (e.g. tr $\rightarrow$ 6.79) and 
the secondary transitions (e.g. 6.79 $\rightarrow$ 0) as well as background lines (identified by their impurity nuclides) are indicated. 
The high-energy part of the spectrum obtained at $E_p = 140$ keV, our lowest bombarding energy, is shown in Fig. \ref{130}: 
the primary transition tr $\rightarrow$ 0 and nearly all secondary transitions are clearly observable.

\section{Experimental procedures, data analyses and results}
\subsection{Excitation energies in $^{15}O$}
\label{excitation}

The energy scale of the Ge detectors used in geometry 2 was determined using the well-known background lines from 
 $^{232}Th$, $^{40}K$ and $^{214}Bi$ and the ground state capture transition at the resonance. 
 The latter energy was calculated from the known 
accelerator energy ($E_R = 259.4\pm0.3$ keV, section \ref{strength}) and the Q value of $^{14}\mathrm{N}(\mathrm{p},\gamma)^{15}\mathrm{O}$ 
($Q = 7296.8\pm 0.5$ keV \cite{Ajz91}), where effects of the Doppler shift and recoil energy were taken into account: $E_x = 7556.4\pm 0.6$ keV. 
The energies of the $\gamma\gamma$ cascades via the 5.18, 6.17 and 6.79 MeV states were then used to 
determine the excitation energies of these states, where the Doppler attenuation factors were left 
as a free parameter in the fit. The results are summarised in table \ref{stati} and are
in good agreement with previous work \cite{Ajz91}.

\subsection{Efficiency and summing effects}
\label{efficiency}
\begin{figure*}
  \includegraphics[width=15cm]{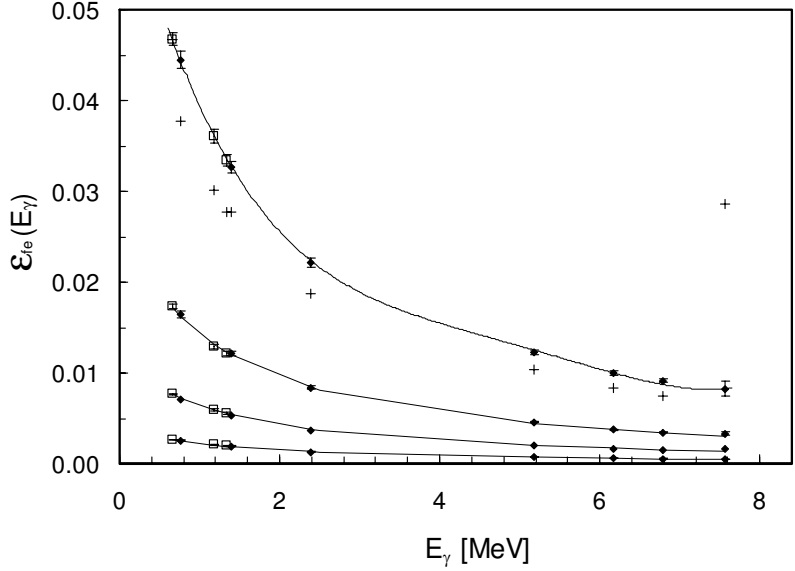}
\caption{Full-energy peak efficiency as function of $\gamma$-ray energy for the geometry 1 
with distances $d = 1.5$, $5.5$, $15.5$, and $20.5$ cm, from top to bottom. The lines through the 
data points are the results from a fit. The crosses are the results omitting the summing effects.}
\label{effi}       
\end{figure*}

Due to (i) the close geometry of the set-up in geometry 1 (Fig. 2), (ii) the cascade structure of the capture process, 
and (iii) the small branching to the ground state, the "summing-in" effect (i.e. the cascade $\gamma$-rays giving 
a contribution to the ground state full-energy peak) and the "summing-out" effect (i.e. a full-energy detection of one cascade  
$\gamma$-ray and any concurrent interaction of the other member of the cascade) have to be taken into account. 
For the efficiency and summing-effect evaluation, we considered  $\gamma$-ray spectra obtained at $55^\circ$
for different distances d from the target (i.e. $d = 1.5$, $5.5$, $10.5$, and $20.5$ cm). 
The spectra involved $^{137}Cs$ and $^{60}Co$ calibrated sources placed at the target position as well as 
the spectra obtained at the $E_R = 259$ keV resonance. The dependence of the full-energy efficiency  $\varepsilon_{fe}$ 
on the  $\gamma$-ray energy $E_\gamma$ and on the distance d were parameterized by the following functions \cite{knol}:
\begin{equation}\label{par1}
\ln{(\varepsilon_{fe})}=a+b\ln({E_\gamma})+c[\ln({E_\gamma})]^2
\end{equation}
\begin{equation}\label{par2}
\varepsilon_{fe}(d) =\frac{1-e^{\frac{d+d_0}{1+\beta\surd{E_\gamma}}}}{(d+d_0)^2}
\end{equation}
In order to take into account the summing-in and summing-out effects, the following expressions were used 
\begin{eqnarray}\label{sum}
Y_{gs}&=&R\bigg(b_{gs} \varepsilon_{fe}(E_{gs})+\sum_i b_i \varepsilon_{fe}(E_i^{sec})\varepsilon_{fe}(E_i^{pri})\bigg) \nonumber\\
Y_{i_{pri}}&=&Rb_i\varepsilon_{fe}(E_{i_{pri}})(1-\varepsilon_{tot}(E_{i_{sec}})) \nonumber\\
Y_{i_{sec}}&=&Rb_i\varepsilon_{fe}(E_{i_{sec}})(1-\varepsilon_{tot}(E_{i_{pri}})) 
\end{eqnarray}
where $Y_{gs}$, $Y_{i_{pri}}$ and $Y_{i_{sec}}$ are the observed yields of the ground state transition and of the 
transitions through the energy level i; the subscripts pri and sec refer to primary and secondary transitions, respectively; 
R is the number of reactions per unit charge, b$_{gs}$ and b$_{i}$ are the branchings, $\varepsilon_{fe}$ and $\varepsilon_{tot}$ 
are the full energy and total efficiencies, respectively. The equation \ref{sum} assumes isotropic 
angular distributions and correlations. 
In these equations the total detection efficiency is parameterized as \cite{knol}:
\begin{equation}\label{par3}
\ln(\frac{\varepsilon_{fe}}{\varepsilon_{tot}})=K_1+K_2\ln(E_\gamma)+K_3(\ln(E_\gamma))^2
\end{equation}
Each excited state that is fed by the resonance then decays to the ground state with a $100\%$ probability: 
this allows to set constraints on the efficiency curve arising from the one-by-one equality of the 
intensities of primary and secondary transitions involving each excited state:
\begin{equation}
\frac{Y_{i_{pri}}}{Y_{i_{sec}}}=\frac{\varepsilon_{fe}(E_{i_{pri}})(1-\varepsilon_{tot}(E_{i_{sec}}))}
{\varepsilon_{fe}(E_{i_{sec}})(1-\varepsilon_{tot}(E_{i_{pri}}))}
\end{equation}
These constraints were used in a global fit to the data. 
The free parameters are the coefficients in the parameterizations (equations \ref{par1}, \ref{par2}
and \ref{par3}) and the four branching ratios of the resonance. 
Fig. \ref{effi} shows the behavior of the efficiency $\varepsilon_{fe}(E)$ versus energy for the four distances:
the data points are those from the reaction (diamonds) and from the sources (squares) and the lines are the fit. 
Points and data refer to distances of $1.5$, $5.5$, $15.5$ and $20.5$ cm starting from the highest curve, respectively.
In order to illustrate the importance of the summing correction, the crosses in the figure represent the values of the efficiency at the closest distance 
which would be obtained neglecting the summing effects: note the large deviation for the ground state capture at 7.5 MeV.
The results shown in Fig. \ref{effi} are consistent with values obtained from Monte-Carlo simulations 
\cite{heide03}.

The resulting branchings of the $E_R = 259$ keV resonance are in good agreement 
with previous work \cite{Ajz91} (table \ref{stati}) except for the ground state 
decay. 

\subsection{Strength, energy and width of the $E_R = 259$ keV resonance}
\label{strength}
The thick-target yield Y$^\infty$ of the resonance is given by the expression \cite{rol88}:
\begin{equation}
Y^\infty = \frac{\lambda^2}{2}\frac{\omega\gamma}{\eta(E)}
\end{equation}
where $\eta(E)\equiv dE/dx$ is the stopping power of the target at the resonance energy. 
For the observed stoichiometry of the TiN target, the compilation \cite{SRIM} leads to 
a stopping power $\eta(E_R) =32.8~eV/(10^{15}atoms/cm^2)$. We find then a resonance 
strength $\omega\gamma=12.9~\pm~0.4$ (stat) $\pm~0.8$ (sys) $meV$. The systematic error 
arises mainly from the uncertainty in the stopping power and the accuracy of the current measurement. 
The strength value is in good agreement with previous work: $\omega\gamma=14~\pm~1$ meV  \cite{Ajz91}
and $\omega\gamma=13.5~\pm~1.2$ meV \cite{ru05}.

\begin{figure*}
  \includegraphics[width=15cm]{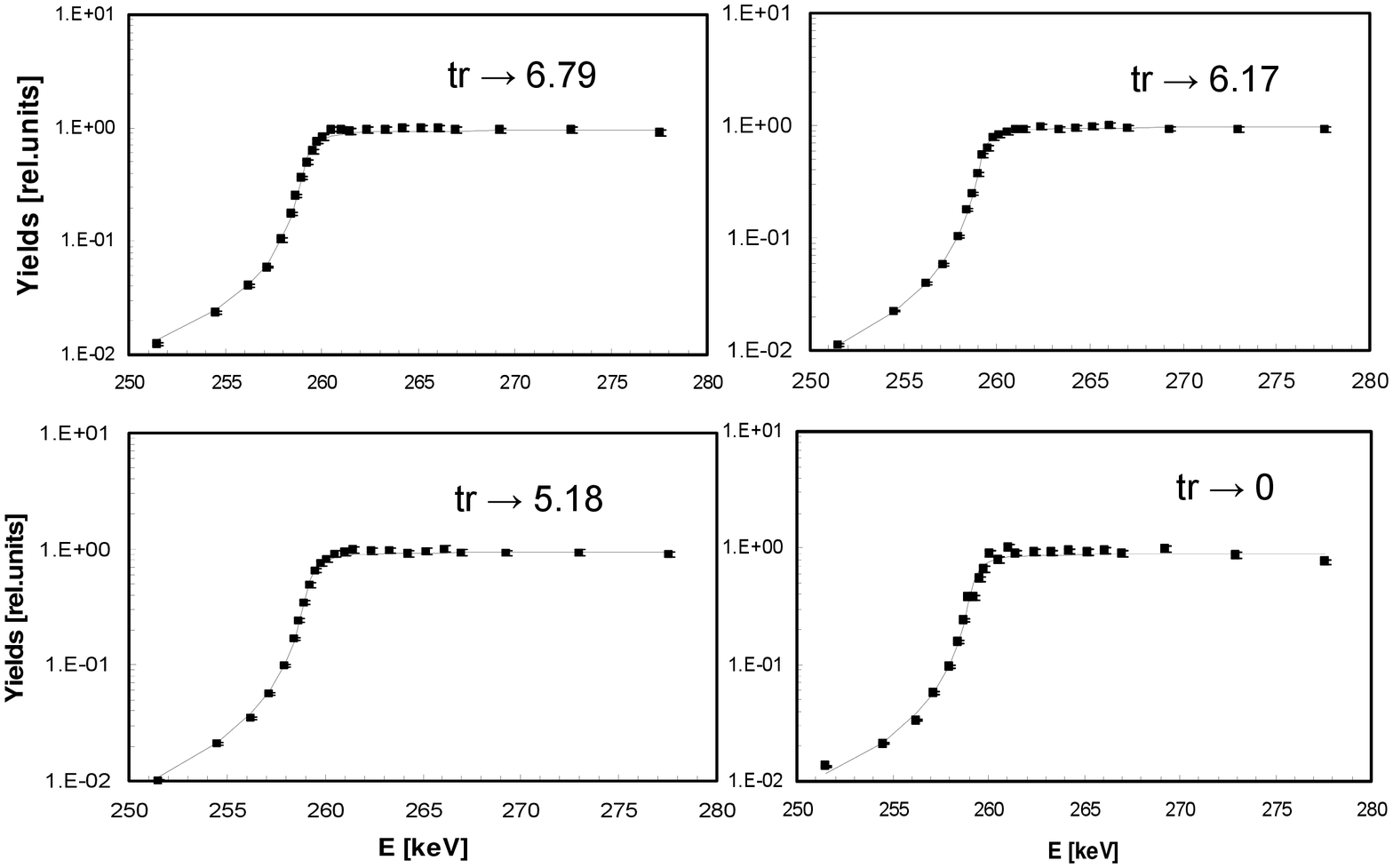}
\caption{Excitation functions of the 4 primary transitions near the $E_R = 259$ keV resonance. 
The lines through the data points are the result of a fit.}
\label{profilo}       
\end{figure*}

In the energy region of the $E_R = 259$ keV resonance, the yield of a capture transition as a function of beam 
energy E is given by the integral:
\begin{equation}
Y(E)=\int_0^E{\frac{\sigma(E)}{\eta(E)}dE}
\end{equation}
With the known stopping power and assuming the cross section as an incoherent sum of a constant non-resonant term plus 
a Breit-Wigner term, we have fitted the experimental curves of all four primary capture 
transitions, where the free parameters were E$_R$ and  $\Gamma_R$. 
The results for all four transitions (Fig. \ref{profilo}) were within their respective errors leading to
$E_R = 259.4\pm 0.3$ keV and 
$\Gamma_R = 0.96\pm 0.05$ keV, 
in good agreement with previous work \cite{schroeder87,Ajz91}. 

\subsection{Angular distributions}
\label{angular}

Angular distributions were measured using geometry 2 at energies E = 206, 313, and 361 keV. Since the J$^\pi$ = 1/2$^+$, 
$E_R = 259$ keV resonance exhibits isotropic distributions, the relative efficiencies of the Ge detectors 
were measured at this resonance. The resulting distributions were fitted with Legendre polynomials
\begin{equation}\label{dist}
W(\theta)=1+a_1Q_1P_1(\theta)+a_2Q_2P_2(\theta) 
\end{equation}
where $a_k$ are the angular distribution coefficients and $Q_k$ the attenuation coefficients calculated according 
to \cite{rose53}. Within experimental uncertainties, all primary and secondary transitions are isotropic \cite{heide03}  
and consistent with the results of Schr\"oder et al. \cite{schroeder87}. As known from \cite{schroeder87}, 
the exception is the primary transition to the 
6.79 MeV state, which exhibits a sin$^2(\theta)$ distribution ($a_2 = -0.50$): $a_2 = -0.79\pm 0.11$, $-0.64\pm 0.06$, 
and $-0.57\pm 0.07$ at E = 206, 313, and 361 keV, respectively, consistent with the data from \cite{schroeder87} at higher energies.

\subsection{Primary peaks}
\label{primary}

Excitation functions have been obtained at proton energies between $E_p = 140$ and $400$ keV for all 
primary transitions. The observed line-shape of a primary transition is determined by the cross 
section behavior $\sigma(E_p)$ in the proton energy interval spanned by the incident beam during 
the slowing-down process in the target, once the transformation from the energy $E_p$ (at which 
the reaction takes place) to the corresponding  $\gamma$-ray energy E$_\gamma$ = E+Q-E$_x$ 
is performed. The shape is also influenced by the energy loss of the protons in the thick target, 
since the stopping power of the protons in TiN is a function of proton energy \cite{And77}. 
Finally, the position of the high-energy rise in the line-shape is influenced by the 
Doppler effect and the recoil of the compound nucleus.

The number of counts N$_i$ in channel i of the $\gamma$-spectrum, corresponding to the energy bin 
E$_{\gamma i}$ to E$_{\gamma i}$ + $\delta$E$_\gamma$ ($\delta$E = dispersion in units of keV per channel) 
is given by the expression
\begin{equation}\label{cts}
N_{i}~=~ \frac{\sigma(E_{pi})\delta
E_{\gamma}\varepsilon_{fe}(E_{\gamma i})b_j }
{\eta(E_pi)}
\end{equation}
for E$_{pi}\leq$ E$_p$ (E$_{pi}$ = proton energy corresponding to channel i, E$_p$ = incident proton energy), 
where $\sigma(E_{pi})$ is the capture cross section, $\varepsilon_{fe}(E_{\gamma i})$ is the $\gamma$-ray 
detection efficiency, and b$_j$ is the branching of the associated decay. The conversion from E$_{\gamma i}$ to E$_{pi}$ 
includes the Doppler and recoil effects. The result is folded with the known detector resolution 
$\Delta$E$_\gamma$ to obtain the experimental line-shape. The spectrum obtained at an energy E$_p$ was 
compared to that obtained with the same target at the E$_{R} = 259$ keV resonance energy: 
the comparison was performed before and after a given run at E$_p$, where the thick-target resonance 
yields Y$^\infty$ were averaged. In this comparison the experimental fit-quantity is the ratio
\begin{equation}\label{ratio}
\frac{N_i}{Y^\infty}=\frac{2\sigma(E_{pi})\delta
E_{\gamma}\varepsilon_{fe}(E_{\gamma i})b_j\eta(E_{Rlab}) }
{\lambda ^2\omega\gamma b_{Rj} \varepsilon_{fe}(E_{R\gamma})\eta(E_{pi})}
\end{equation}
The energy dependence of the full-energy efficiency is known; the summing-out correction for the cascade 
transitions via the three excited states is smaller than 1$\%$; the summing-in correction for the capture to 
the ground state depends on $b_j$ and $b_{Rj}$. Thus, for the ground state capture a correction has 
to be applied which can be obtained replacing equation \ref{ratio} with:
\begin{equation}\label{ratiosum}
\frac{N_{gs}}{Y_{gs}^\infty}\propto 
\frac{\sigma_{gs}(E_{p})\varepsilon_{fe}(E_\gamma^{gs})
+\sum_i{\sigma_i(E_{p})\varepsilon_{fe}(E_{\gamma i}^{sec})\varepsilon_{fe}(E_{\gamma i}^{pri})}}
{\sigma_{gs}(E_{Rp})\varepsilon_{fe}(E_{\gamma R}^{gs})
+\sum_i{\sigma_i(E_{Rp})\varepsilon_{fe}(E_{\gamma R i}^{sec})
\varepsilon_{fe}(E_{\gamma R i}^{pri})}} 
\end{equation}
 using $\sigma_i(E_p)$ deduced from equation \ref{ratio}.
 
The summing-in correction for the ground state transition is not negligible and becomes increasingly uncertain 
with decreasing proton energy. This was the main reason why we stopped the measurements at E$_p = 140$ keV 
although from a statistical point of view and background considerations we could go lower 
in energy (Fig. \ref{130}). In the case of the other capture transitions the summing-out correction depends only 
on the precision of the efficiency curve.

\begin{figure*}
  \includegraphics[width=15cm]{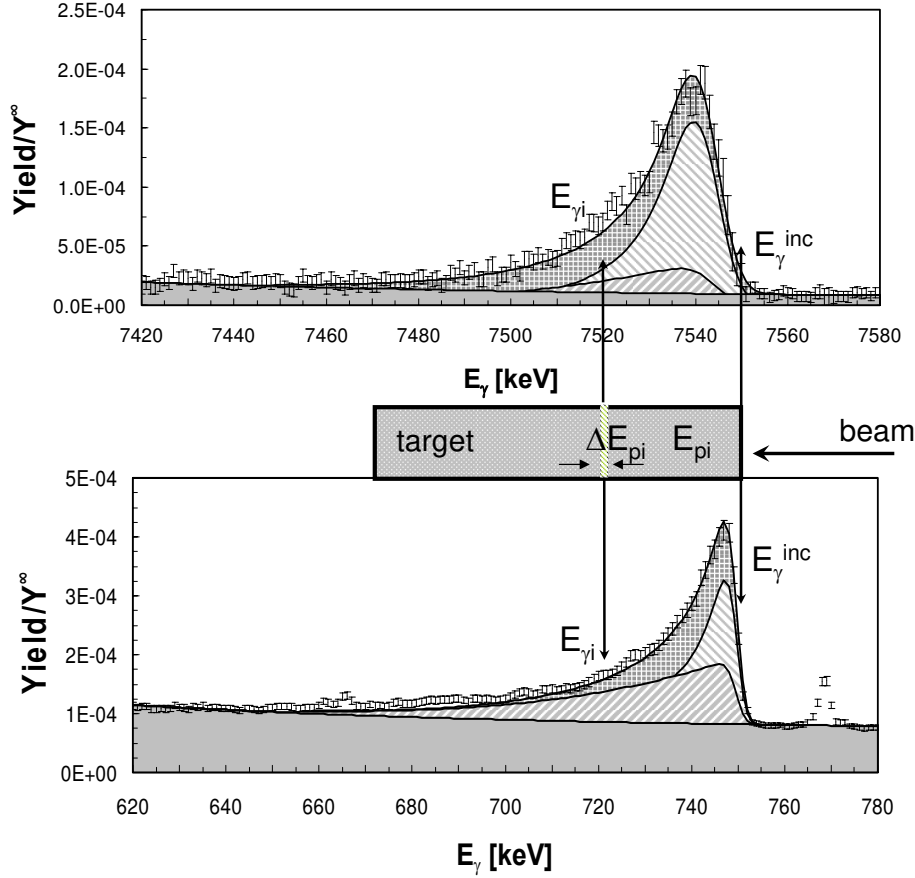}
\caption{Gamma-ray spectra near the ground state transition (top) and the primary 
transition to the 6.79 MeV state obtained at E$_p = 250$ keV (for details, see text).}
\label{fit}       
\end{figure*}

The last step in the analysis of the primary peaks is to introduce the function for the cross section 
in equation \ref{ratio}. Since we have analysed the cross section above and below the 
E$_R = 259$ keV resonance, we have written - in first approximation - the cross section as an 
incoherent sum of a non-resonant term, 
which assumed a constant astrophysical S(E) factor S$_{NR}$, and a resonant term described by the 
Breit-Wigner formula:
\begin{equation}\label{cross}
\sigma(E_{p_i})=\frac{S_{NR}e^{-2\pi\eta}}{E_{p_i}}+
\frac{\lambda^2}{\pi}\omega\gamma\frac{\Gamma}{(E_{p_i}-E_R)^2+(\Gamma/2)^2)}
\end{equation}
These expressions were fitted to the $\gamma$-ray spectrum yields, superimposed on an exponential background. 

\begin{figure*}
  \includegraphics[width=15cm]{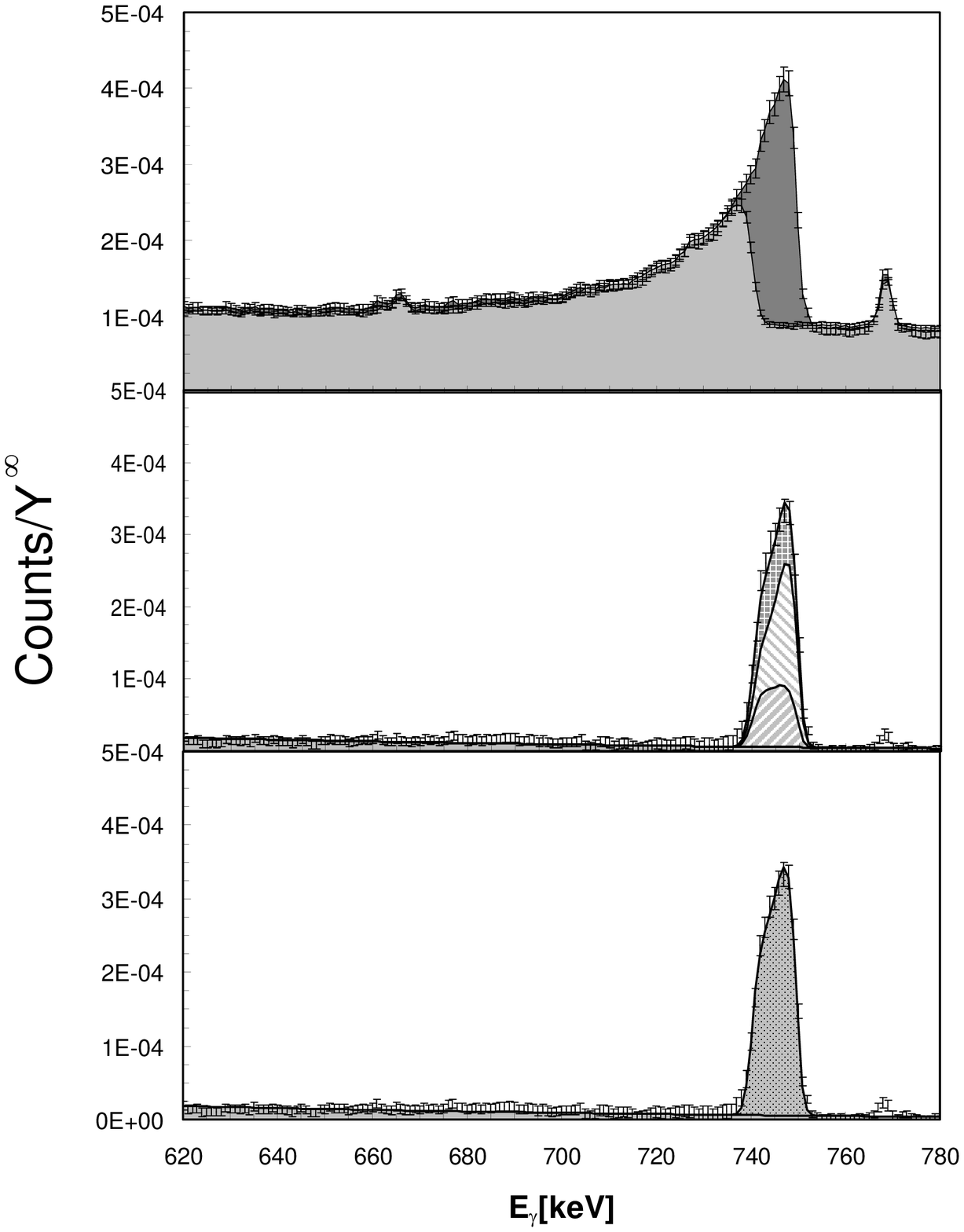}
\caption{Overlay of spectra near the primary transition to the 6.79 MeV 
state obtained at E$_p = 250$ and 240 keV (top) and the difference of both spectra (center).
Bottom part shows fit using the R-matrix analysis result.}
\label{picchi}       
\end{figure*}

An example of this fitting procedure is given in Fig. \ref{fit}, where the $\gamma$-spectra in the region of the ground state 
transition (top figure) and of the primary transition to the 6.79 MeV state (bottom figure) are shown as 
obtained at E$_p = 250$ keV: the gray area is the background, the left-oriented stripes area and the right-oriented
stripes area represent the non-resonant and resonant parts of the cross section, respectively, and the squared area 
is their sum. In the fits we used as free parameters the non resonant astrophysical factor and the background 
parameters. After the 
R-matrix analysis (section \ref{R-matrix}) this procedure has been checked resulting in a fit shown
at the bottom of Fig. \ref{picchi} for the transition to the 6.79 MeV state. 

Since the non-resonant astrophysical S(E) factor can be considered as constant only in a short energy range, 
our thick target may represent a too large energy window. Indeed the value, where the fitting value 
converges, varies with energy. Moreover, in some cases background peaks can be present in the line-shape 
region, due either to natural background or beam-induced background. For example, in the region of 
the ground state peak there are contamination peaks due to the presence of $^{18}$O in the target, and 
in the region of the low-energy primaries there are many lines due to the natural radioactivity of 
the rocks. Thus, an alternative way to analyse the data, i.e. simulating a thin target, is to subtract 
two spectra acquired at different energies. The resulting spectrum is equivalent to one acquired at 
the higher energy with a target thickness equal to the difference in incident energy 
($\Delta$ = E$_{max}$ - E$_{min}$). 
An example is shown in Fig. \ref{picchi}: the upper part shows the 
$\gamma$-spectra obtained at E$_p = 250$ keV (dark gray area) and 240 keV (light gray area) for 
the primary transition to the 6.79 MeV state; the center part shows the $\gamma$-spectrum resulting 
from the subtraction of both spectra, which is equivalent to a spectrum acquired at an energy of 
250 keV with a target of $\Delta = 10$ keV thickness. 
After the final R-matrix analysis was completed we have reinvestigated this differential method,
using the R-matrix results, which is shown in the bottom part of the Fig \ref{picchi}. Excellent agreement 
with the approximation shown in the upper and center parts is noted.
The differential method in combination with the line-shape analysis has been used for the 
primary transitions.

\subsection{Secondary transitions}

The analyses of the secondary transitions have several advantages: 
(i) since the  $\gamma$-energies of all secondary transitions are above 5 MeV, 
we can make full use of the advantage of an underground laboratory; 
(ii) the expected peaks are relatively narrow (compared to the primaries) and hence 
the peak to background ratio is improved; 
(iii) there is no efficiency correction necessary when comparing the off-resonance 
yield with that on-resonance; 
(iv) the interpretation of the observed yield in combination with an 80 keV thick 
target is nearly identical with the integral of the cross section from E=0 to the 
incident energy, thus target-profile corrections are not necessary; 
(v) angular distribution effects are expected to be negligible (section \ref{angular}).

The yields of all secondary transitions were obtained by fitting the background with 
a linear function at the low-energy and high-energy sides of the peak which was then 
subtracted from the total peak intensity. The yield was then compared with the 
corresponding thick-target yield step Y$^\infty$ at the resonance. The resulting thick-target 
yield curves are shown in the inserts of Figs. \ref{6.79}, \ref{5.18} and \ref{6.17} for the secondary transitions 
E$_\gamma$  = 5.18, 6.17, and 6.79 MeV, respectively. Each yield ratio corresponds to the integral 
from E = 0 to the respective incident beam energy E$_i$ taken in the center of mass system. 
Equation \ref{ratio} can thus be written:
\begin{equation}\label{ratiosec}
\frac{Y(E_i)}{Y^\infty}=\frac{2}
{\lambda^2\omega\gamma}
\int_0^{E_{i}}{\frac{\sigma(E)b_j\eta(E_R)}{\eta(E)}dE}
\end{equation}

where $\omega\gamma$ are the respective values for the individual transitions of the E$_R$ = 259 keV resonance. 
These yield ratios are corrected for summing-out effects. 
Note that the stopping power only enters equation \ref{ratiosec} as a ratio and has not 
to be known absolutely.
In principle one could also use the differential method (subtraction of the secondary peak contents from 
two runs with successive beam energies) to determine the energy behavior of the cross section.
However, the statistical error increases by subtracting two large numbers together with 
the uncertainty in the determination of the effective beam energy. Since
the energy dependence of the cross section is in principle known from the R-matrix analysis  of the primaries
we have adopted the method of fitting the thick target yield curve (equation \ref{ratiosec}) using the 
R-matrix analysis with a fine tuning of the fit parameters. The solid curves through the data 
points of the secondary transitions, the insert in Figs. \ref{6.79}, \ref{5.18} and \ref{6.17},
are the results of integrated R-matrix calculation.

To complete the study of the secondary peaks we have studied the 5.24 MeV (see fig. \ref{5.24})transition. 
The 5.24 MeV state is 
populated by cascades through the 7.28 and 6.85 MeV states, which in turn decay 100 \% to the 
5.24 MeV state \cite{schroeder87}. 
In the 259 keV resonance the 5.24 MeV state is populated directly with 0.6 $\%$ branching.
At energies below this energy the primary peaks feeding the 7.28 and 6.85 MeV states were not visible
in our spectra.

\begin{figure*}
  \includegraphics[width=15cm]{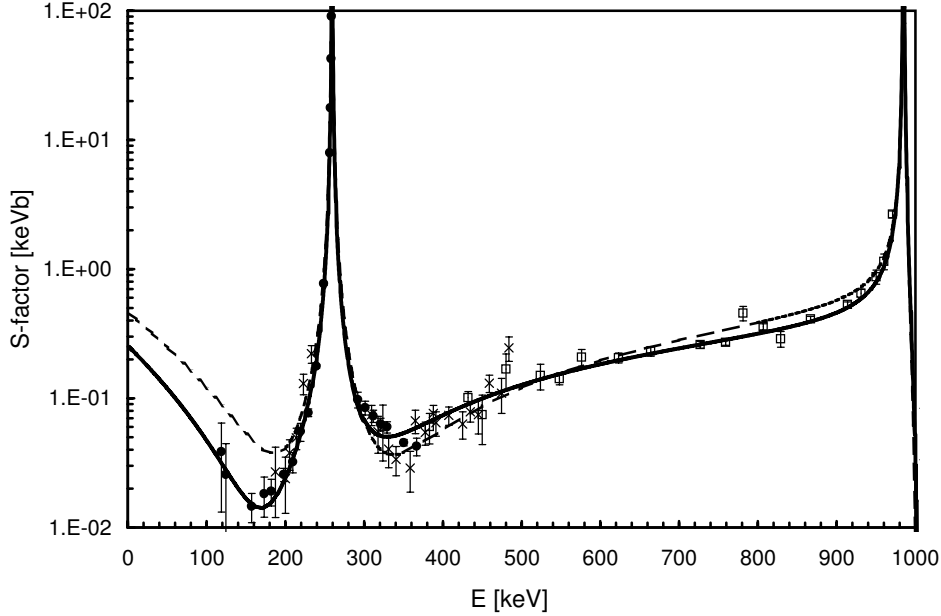}
\caption{The astrophysical S$_{gs}$(E) factor for the capture to the ground state 
is shown from the LUNA studies (filled-in points), from previous work \cite{schroeder87} (open squares) and 
from the LENA group \cite{ru05} (crosses). 
The thick line through the data points represents our R-matrix fit, while the dashed line represents the LENA R-matrix fit.} \label{gs}
\end{figure*}

\begin{figure*}
  \includegraphics[width=15cm]{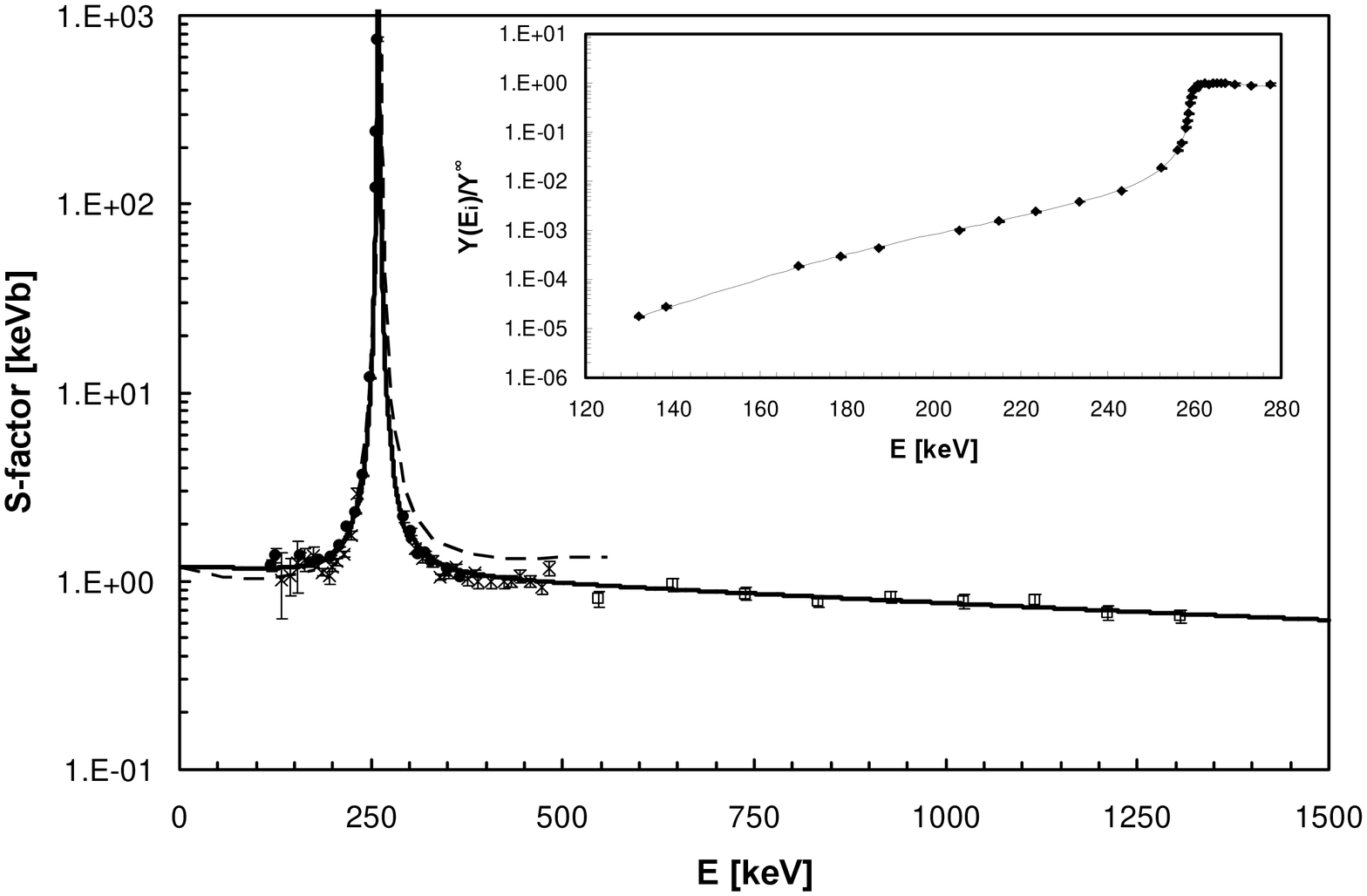}
\caption{The astrophysical S$_{6.79}$(E) factor for the capture to the 6.79 MeV state is 
shown from the LUNA studies (filled-in points), recent work \cite{klug05} (open squares)and 
from the LENA group \cite{ru05} (crosses). 
The insert shows the thick-target yield data for the 6.79 MeV secondary transition. The lines through the 
data points represent the results of our R-matrix fit. The dashed curve is scanned from \cite{TUNL03}.} \label{6.79}
\end{figure*}

\begin{figure*}
  \includegraphics[width=15cm]{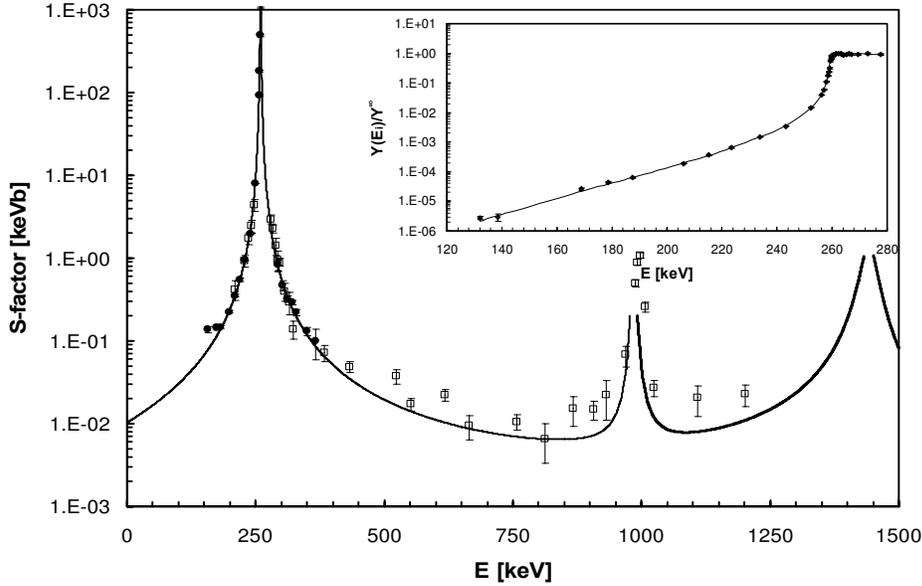}
\caption{The astrophysical S$_{5.18}$(E) factor for the capture to the 5.18 MeV state is shown 
from the LUNA studies (filled-in points) and previous work \cite{schroeder87} (open squares). 
The insert shows the thick-target yield data for the 5.18 MeV secondary transition. 
The lines through the data points represent the results of our R-matrix fit.} \label{5.18}
\end{figure*}

\begin{figure*}
  \includegraphics[width=15cm]{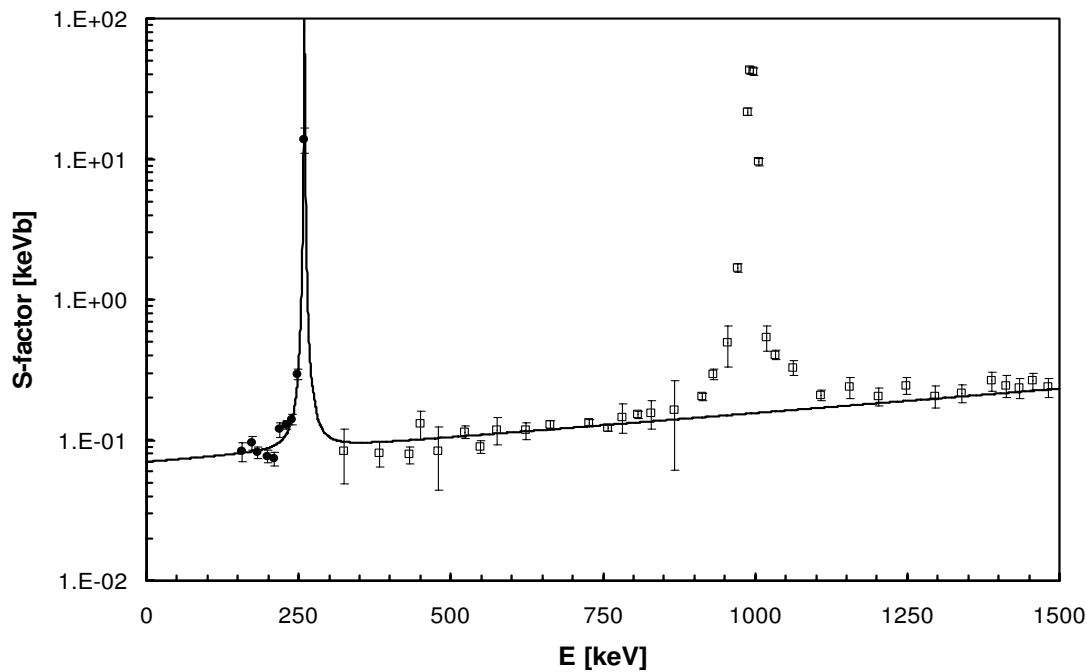}
\caption{The astrophysical S$_{5.24}$(E) factor for the capture to the 5.24 MeV state (only secondary transition was observed) is shown from the LUNA 
studies (filled-in points) and previous work \cite{schroeder87} (open squares). 
The line through the data points represents the energy dependence of the DC process including the 259 keV
resonance contribution.} \label{5.24}
\end{figure*}

\begin{figure*}
  \includegraphics[width=15cm]{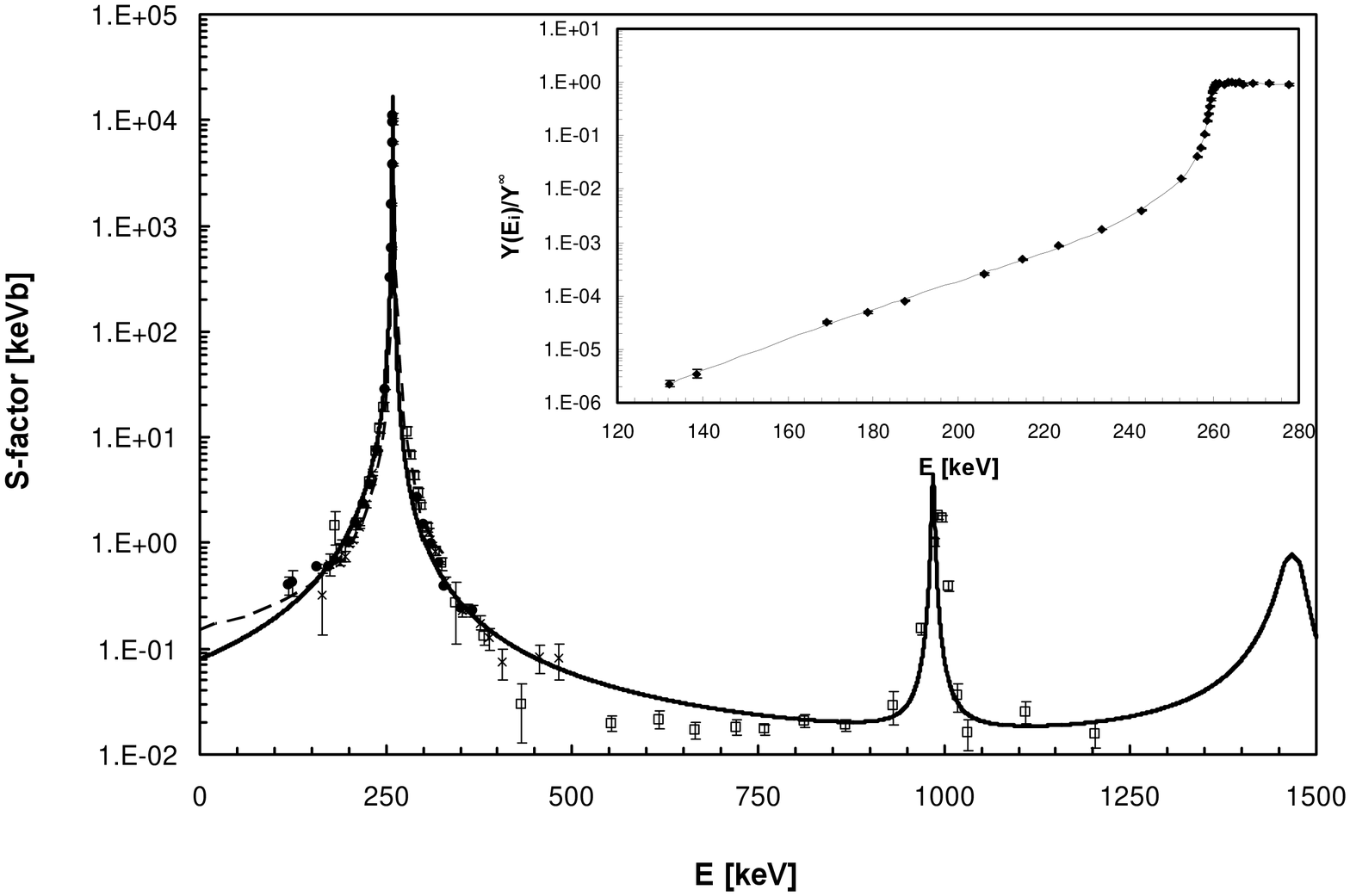}
\caption{The astrophysical S$_{6.17}$(E) factor for the capture to the 6.17 MeV state is shown 
from the LUNA studies (filled-in points), previous work \cite{schroeder87} (open squares) and 
from the LENA group \cite{ru05} (crosses). 
The insert shows the thick-target yield data for the 6.17 MeV secondary transition. 
The lines through the data points represent the results of our R-matrix fit.
The dashed curve is scanned from \cite{TUNL03}.} \label{6.17}
\end{figure*}

The results of both analyses are shown in figures \ref{gs}-\ref{6.17} and numerical values are 
given in table \ref{results}.
The solid curves in the insert of figures \ref{gs}-\ref{6.17} are the results of integrated R-matrix calculations. 

\section{R-matrix analysis and discussion}
\label{R-matrix}
The experimental results for the transition to the ground state and to the E$_x$ = 6.79 MeV state were reported in \cite{formicola04} 
and an R-matrix \cite{lane58,angulo01} analysis was described for these two transitions with the results: 
S$_{gs}(0)=0.25\pm 0.06$ keVb and $S_{6.79}=1.35\pm0.05$ keVb, where the quoted errors include
only statistical uncertainties of the data; this error quotation was also applied below.
The cross section for all transitions to excited states can be determined via the primary and secondary 
transitions. The S(E) determination of the primary transitions has been described in section \ref{primary}
and can be directly used for an R-matrix analysis. The peak of a secondary transition for an infinitely 
thick target contains the integrated yield from zero energy to the respective beam energy. 
Thus, its interpretation requires knowledge of the energy dependence of the cross section. 
This energy dependence is provided by the R-matrix fit on the primary data as a starting 
condition with a subsequent fine tuning of the fitting parameters. 
Such a procedure was used for the transitions to the 5.18, 6.17 and 6.79 MeV states. 
The best choice of the channel radius for fitting the transition to the ground state 
and 6.79 MeV state was found \cite{formicola04} to be a = 5.5 fm, 
which was used for the present R-matrix calculations.

\subsection{Transition to the ground state}

The data from LUNA and previous works for the ground state capture are shown in Fig.\ref{gs}.
We refer to the analysis described in our previous paper \cite{formicola04}.
Briefly, the data were fitted including the $3/2^+$ sub-threshold state, the $1/2^+$, 259 keV, 
the $3/2^+$, 987 keV and the $3/2^+$, 2187 keV resonances as well as a 
background pole located at 6 MeV. 
For the sub-threshold state, we used the reduced width obtained from the fit of the 
data for the 6.79 MeV transition.
The fit parameters were the $\Gamma_\gamma$ of the subthreshold state, the 987 keV and 2187 keV 
resonances, the $\Gamma_p$ and $\Gamma_\gamma$ of the background pole.
For the external contribution we used the ANC of \cite{mukhamedzhanov03} as a starting value.
The corresponding $\Gamma_\gamma$ value for the subthreshold state is $0.8 \pm 0.4$ $eV$, which is
in good agreement with the value from a life time measurement by \cite{bertone01} 
$\Gamma_\gamma=0.41^{+0.34}_{-0.13}$ $eV$ as well as with $\Gamma_\gamma=0.95^{+0.60}_{-0.95}$,
the value from coulomb excitation work \cite{yamada04}.
The minimum in the S-factor shape at energy lower than 200 keV is due to an interference effect
between the sub-threshold state and the nonresonant capture mechanism. The extrapolation of
the S-factor is based on the energy dependence of the complete data set and not on the lowest
data points. The extrapolated value of the ground state S-factor is S$_{gs}(0)=0.25\pm 0.06$ keVb. 
Since the publication of \cite{formicola04} a new experiment was reported \cite{ru05} with an 
extrapolated S$_{gs}(0)=0.49\pm 0.08$ keVb (table \ref{S0}). 
The present data and those of \cite{ru05} are in excellent agreement within their respective errors 
but the extrapolated S$_{gs}(0)$ 
differ by a factor of two however still within their 2$\sigma$ errors (Fig. \ref{gs}). 
The major difference in both R-matrix analyses is that 
in \cite{formicola04} the high energy data of \cite{schroeder87} - corrected for summing-in effects - 
were included in our analysis forcing the S$_{gs}$(E) to be higher above the 259 keV resonance and 
lower below the resonance. 

\subsection{Transition to the E$_x$=6.79 MeV state}

It was found that the  $\chi^2$ of the R-matrix fit for the transition to E$_x=6.79$ MeV state was 
primarily determined by the data of \cite{schroeder87} forcing the fit to be higher than the lowest 
energy data points of \cite{formicola04}. 
It was thus desirable to check the data above the $E_R=259$ keV resonance. 
This experiment has been performed recently at the 4 MV Dynamitron tandem accelerator in Bochum over the
energy range E$_p$ = 0.6 to 1.3 MeV \cite{klug05} using the setup of geometry 1 (Fig. \ref{geom1}). 
The results are used together with the data of \cite{formicola04} in a new R-matrix fit 
and are shown in fig. \ref{6.79}. 
The high-energy data are lower than those of \cite{schroeder87} but still within the respective 2$\sigma$ errors. 
In this case the fit parameters for the R-matrix analysis 
were $\Gamma_\gamma$ of the $E_R=259$ keV resonance 
and the ANC of the 6.79 MeV state. 
The resulting fit of the thick target yield curve for the $E_\gamma=6.79$ MeV secondary peak is also 
shown as insert in fig.\ref{6.79}.
The new extrapolated value S$_{6.79}(0)=1.21\pm0.05$ keVb is within 2$\sigma$ of the value given in 
\cite{formicola04} and in excellent agreement 
with S$_{6.79}(0)=1.15\pm0.05$ keVb from \cite{ru05}. 
The new ANC value of $C=4.7\pm0.1 $ fm$^{-1/2}$ is still in excellent agreement with the ANC results 
\cite{mukhamedzhanov03} 
($C = 5.2\pm 0.7 $ fm$^{-1/2}$) and  \cite{bertone02} ($C = 4.6\pm 0.5 $ fm$^{-1/2}$). 
Also shown in fig.\ref{6.79} is the R-matrix fit of \cite{TUNL03} as a dashed curve: 
the fit does not reproduce the present data. 

\subsection{Transition to the E$_x$=5.18 MeV state}

This transition can be fitted with the inclusion of the two $J^\pi=1/2^+$ resonances at E$_R$ = 259 
and 1446 keV and the $J^\pi=3/2^+$ resonance at 
E$_R$ = 987 keV, which is added incoherently. 
The result is shown in fig. \ref{5.18} for the primary and 
secondary transitions. A small external DC contribution could be considered 
but has been neglected here since it amounts 
to less than 1$\%$ to the total S-factor. 
The resulting S$_{5.18}(0)=0.010\pm0.003$ keVb can be compared with $0.018\pm0.003$ from \cite{schroeder87}. 

\subsection{Transition to the E$_x$=5.24 MeV state}

The E$_\gamma =5.24$ MeV secondary transition was observed with a similar low intensity as the 
E$_\gamma =5.18$ MeV transition. 
Due to the lack of detailed data, a R-matrix fit was not attempted and only a fit of 
an exponential representing the external DC contribution was used (Fig. \ref{5.24}) 
yielding S$_{5.24}(0)=0.070\pm0.003$ keVb compared to S$_{5.24}(0)=0.064\pm0.002$  keVb 
from \cite{schroeder87}.

\subsection{Transition to the E$_x$=6.17 MeV state}

The fitting of the transition to the 6.17 MeV state (Fig. \ref{6.17}) requires an external DC 
contribution together with the two $J^\pi=1/2^+$ resonances at E$_R = 259$ and 1446 keV 
which all interfere with each other, together with the $J^\pi=3/2^+$ resonance at 
E$_R = 987$ keV added incoherently (Fig. \ref{structure}). 
Again, the data above E$_p = 350$ keV are from \cite{schroeder87}. 
The best fit was obtained with the respective resonant $\Gamma_\gamma$ widths 22.6, 70, and 10 meV. 
In particular, the fit at energies between the two $J^\pi=1/2^+$ resonances is not very 
sensitive to the choice of the ANC but it is sensitive to this choice at the low-energy 
wing of the E$_R = 259$ keV resonance. 
The yield of the secondary transition (insert in Fig. \ref{6.17}) is fitted best with 
$C=0.2\pm0.1 $ fm$^{-1/2}$ but is not acceptable if C exceeds 0.3 fm$^{-1/2}$. 
The result is within two standard deviation of \cite{mukhamedzhanov03} $C=0.47\pm0.03 $ fm$^{-1/2}$ and 
of \cite{bertone02} $C=0.45\pm0.05 $ fm$^{-1/2}$ after conversion to the present coupling scheme.
The extrapolated values $S_{6.17}(0)=0.14\pm0.02$ \cite{schroeder87}  
and $0.13\pm0.02$ keVb \cite{mukhamedzhanov03} are consistent with the 
present result  $S_{6.17}(0)=0.08\pm0.03$ keV b. 
Recently, the analysing power was measured \cite{TUNL03} at E$_p$ = 270 keV; 
the result together with the data of \cite{schroeder87} led to $S_{6.17}(0)=0.16$ keV b, 
which is a factor two higher than our value. The agreement of their fit (scanned from \cite{TUNL03}) 
with our data at the low-energy wing of the 259 keV resonance is quite good but seems to 
become too high at the high-energy side of the 259 keV resonance where the influence of the 
M1 component increases. Unfortunately the authors \cite{TUNL03} only show their 
fit up to E$_p = 330$ keV so that further conclusions are not possible. 
The analysis of \cite{ru05} with S$_{6.17}(0)=0.04\pm0.01$ keVb is a factor two 
lower than the present result. Our data extend towards lower energies than those 
of \cite{ru05} and in our R-matrix analysis an external DC contribution was included 
which interferes with both $J^\pi=1/2^+$ resonances; 
thus, no background pole needs to be considered. 
Taking all extrapolations just discussed into account, the spread in S$_{6.17}(0)$ 
is of the order of $\pm$ 0.06 keVb which corresponds to 4$\%$ of the total S(0)-factor.

\subsection{Total S-factor}

Our present total S-factor, based primarily on R-matrix fits yields 
S$_{tot}(0)=1.61\pm0.08$ keVb. This result is lower by 6$\%$ than the value given 
in \cite{formicola04} mainly due to the revised analysis of the transition to 
the 6.79 MeV state. The present result is in good agreement with \cite{mukhamedzhanov03,ru05} 
but differs in the weight of the contributions from the various transitions (table \ref{S0}).

\subsection{Reaction rates}

We obtained an improved extrapolation for S$_{tot}$(0) (table \ref{S0}) on the basis of the present 
R-matrix analysis where all presently available experimental information were considered within their 
two $\sigma$ limits. The total reaction rate for the $^{14}\mathrm{N}(\mathrm{p},\gamma)^{15}\mathrm{O}$ 
reaction was calculated numerically from the R-matrix results with \cite{NACRE}:
\begin{eqnarray}
N_A<\sigma v>&=&3.7313\cdot 10^{7}\mu^{-\frac{1}{2}}T_9^{-\frac{3}{2}}\cdot \nonumber \\
&&\int_0^\infty{S(E)e^{(-2\pi\eta)}e^{(11.605E/T_9)}dE}
\label{reactrate}
\end{eqnarray}
in units of $cm^3mole^{-1}s^{-1}$, where S(E) is in MeVb and E in MeV and $\mu$ is the reduced mass. 
The resulting rates can be expressed within 0.5$\%$ using the following analytic expression for T$_9 <$ 2:
\begin{eqnarray}
N_A<\sigma v>&=&a_1 10^7 T_9 ^{-2/3}\exp{[(a_2T_9^{-1/3}-(T_9/a_3)^2)]}\cdot \nonumber \\
             &&(a_4+a_5T_9+a_6T_9^2+a_7T_9^3)+                               \nonumber \\
             &&a_8 10^3T_9^{-3/2}\exp{[a_9T_9^{-1}]}+                        \nonumber \\
             &&a_{10} 10^2T_9^{a_{11}}\exp{[a_{12} T_9^{-1}]}
\label{analytic}
\end{eqnarray}
where:
\begin{center}
\begin{tabular}{ lll }
\hline
$a_1   = 3.12    $   &   $a_5   =  -1.50  $  &  $a_9      =  -2.998 $              \\
$a_2   = -15.193 $   &   $a_6   =  17.97  $  &  $a_{10}   =  8.42   $              \\
$a_3   = 0.486   $   &   $a_7   =  -3.32  $  &  $a_{11}   =  0.0682  $              \\
$a_4   = 0.782   $   &   $a_8   =  2.11   $  &  $a_{12}   =  -4.891  $              \\
\hline
\end{tabular}
\end{center}
\begin{center}
\begin{tabular}{ lll }
\hline
$a^{low}_1   = 2.76    $   &   $a^{low}_5   =  -1.40  $  &  $a^{low}_9      =  -2.998 $              \\
$a^{low}_2   = -15.193 $   &   $a^{low}_6   =  15.82  $  &  $a^{low}_{10}   =  8.44   $              \\
$a^{low}_3   = 0.503   $   &   $a^{low}_7   =  -3.32  $  &  $a^{low}_{11}   =  0.0682 $              \\
$a^{low}_4   = 0.804   $   &   $a^{low}_8   =  2.03   $  &  $a^{low}_{12}   =  -4.987 $              \\
\hline
\end{tabular}
\end{center}
\begin{center}
\begin{tabular}{ lll }
\hline
$a^{high}_1   = 3.44    $   &   $a^{high}_5   =  -1.59  $  &  $a^{high}_9      =  -2.997  $              \\
$a^{high}_2   = -15.193 $   &   $a^{high}_6   =  19.83  $  &  $a^{high}_{10}   =  8.42    $              \\
$a^{high}_3   = 0.475   $   &   $a^{high}_7   =  -3.30  $  &  $a^{high}_{11}   =  0.0681  $              \\
$a^{high}_4   = 0.771   $   &   $a^{high}_8   =  2.18   $  &  $a^{high}_{12}   =  -4.807  $              \\
\hline
\end{tabular}
\end{center}

\begin{figure*}
  \includegraphics[width=15cm]{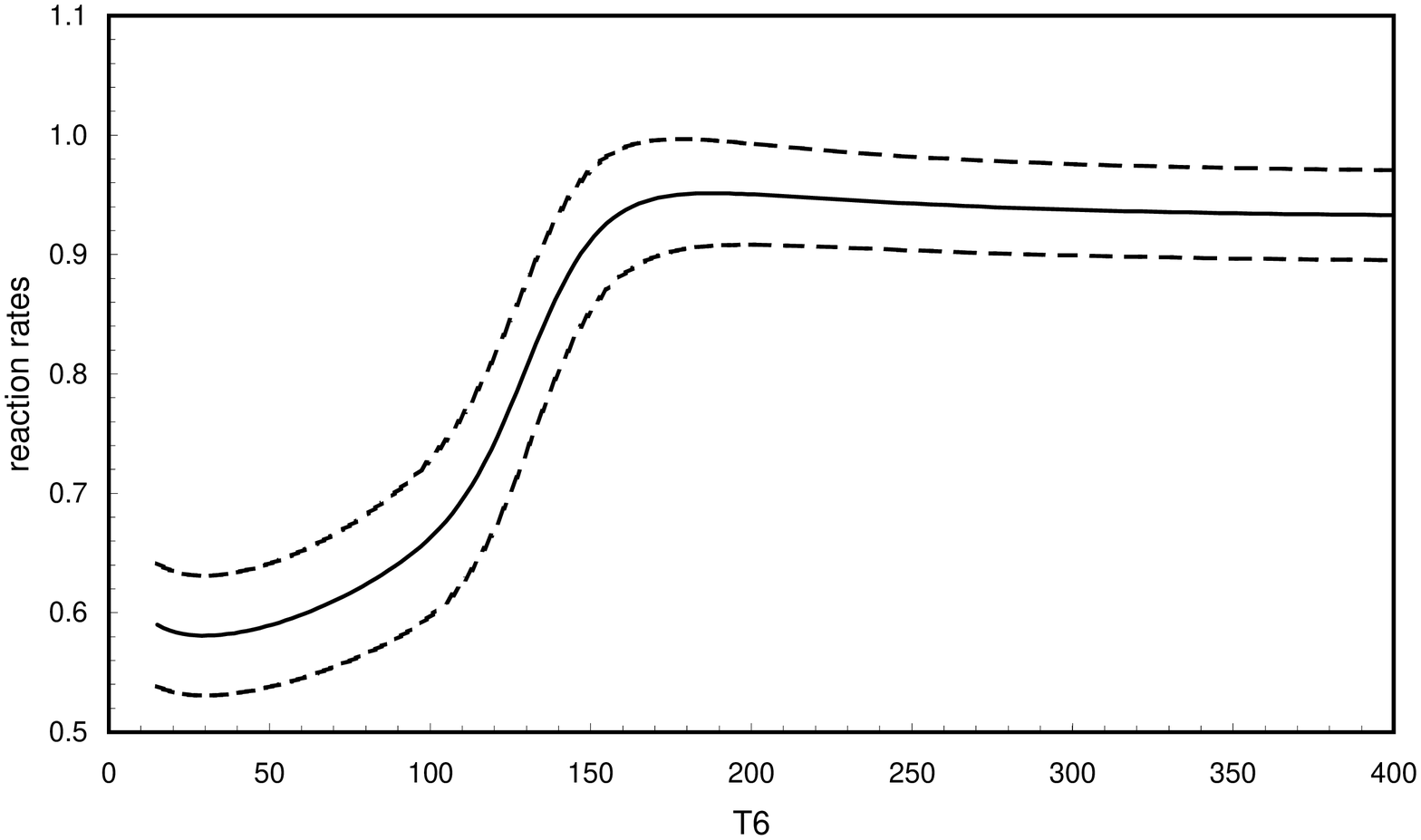}
\caption{Reaction rate from present work is compared with that of the NACRE compilation 
\cite{NACRE}. The dashed curves represent the uncertainty of the present reaction rate.} \label{rate}
\end{figure*}

The parameters labeled "low" and "high" in equation \ref{analytic} correspond to the two $\sigma$ limits 
of present S$_{tot}$(0) extrapolation and include the error in the strength determination $\omega\gamma$ 
of the 259 keV resonance. The results are compared in fig. \ref{rate} with the rates given in the NACRE 
compilation \cite{NACRE}. They confirm the conclusion of \cite{angulo01} that the rate has to be reduced 
by nearly a factor of two at low temperatures, but it is in good agreement with NACRE \cite{NACRE} above 
T$_6$=150. 

It can be also concluded from the present analysis that the data above the 259 keV resonance are of crucial 
importance for a reliable extrapolation. This finding emphasizes that one experiment alone cannot solve 
the problem at low temperatures and detailed analysis of the nuclear structure of $^{15}O$ is required.  

A recent experimental determination of the total S-factor at very low energies down to E=70 keV 
\cite{gasta05} is in good agreement with the present R-matrix calculations. 
While extrapolation of these data to lower energies requires a detailed knowledge of the energy 
dependence of the various contributions to the total S-factor, this experiment \cite{gasta05} gives experimental 
certainty of the reaction rate better then 15\%  for T$_6$$>$90 without any extrapolation procedure. 
This clearly represents a major improvement in the evaluation of the reaction rate for this temperature regime. 

In conclusion, with the present determination of the reaction rates we confirm the astrophysical
consequences in the determination of the age of the Globular Clusters quoted in \cite{imbriani04}, and in the 
CNO solar neutrino fluxes \cite{bp04,d04}.

\section*{Acknowledgments} 
We are grateful to 
to the technical staff of National Laboratory of the Gran Sasso.
Two of the authors, C.A. and H-P.T., gratefully acknowledge A. Champagne for fruitful discussions.
Some of us participate in the PRIN-2004 grant. 
This work was partially supported by the Belgian Inter-University Attraction Poles under project number P5/07,
by FEDER - POCTI/FNU/41097/2001, OTKA T42733 and EU contract RII-CT-2004-506222.


\begin{table*} 
\begin{center}
\begin{tabular}{ c c  }
\hline
  \hline
  E$_X$ [{\rm keV}] & Branching [$\%$] \\
  
\begin{tabular}{ l c }\hline present work~~~~~~~~&\cite{Ajz91}~~~~\\ \hline\end{tabular}&\begin{tabular}{ c c c} 
\hline present work&~~~~~~~~\cite{Ajz91}~~~~~~~~~~&\cite{ru05}~~~~~~~~~~\\ \hline\end{tabular}\\

\begin{tabular}{ c c } $5180.8\pm0.2$& $5183.0\pm1.0$\end{tabular}             & \begin{tabular}{ l c c} $17.1\pm0.2$& $~~~~15.8\pm0.5$    & $17.3\pm0.2$    \end{tabular}\\ 
\begin{tabular}{ c c } $~~~~~~~~~~~~~~~~$& $5240.9\pm0.3$\end{tabular}         & \begin{tabular}{ l c c} $~0.6\pm0.3$& $~~~~~~~~~~~~~~~~$  & $~~~~~~~~~~~~$  \end{tabular}\\ 
\begin{tabular}{ c c } $6172.3\pm0.2$& $6173.3\pm1.7$\end{tabular}             & \begin{tabular}{ l c c} $57.8\pm0.3$& $~~~~57.5\pm0.6$    & $58.3\pm0.5$    \end{tabular}\\ 
\begin{tabular}{ c c } $6791.7\pm0.2$& $6793.3\pm1.7$\end{tabular}             & \begin{tabular}{ l c c} $22.9\pm0.3$& $~~~~23.2\pm0.4$    & $22.7\pm0.3$    \end{tabular}\\ 
\begin{tabular}{ c c } $7556.4\pm0.6$& $7556.5\pm0.4$\\ \hline\end{tabular}    & \begin{tabular}{ l c c} $~~~1.6\pm0.1$& $~~~~~3.5\pm0.6$    & $~1.70\pm0.07$    \\ \hline\end{tabular} \\ 
\\ \hline
\end{tabular}
\caption{Excitation energies of $^{15}$O states and low energy at the E$_R$ = 259 keV resonance.}
\label{stati}
\end{center}
\end{table*}

\begin{table*}
\begin{center}
\begin{tabular}{c c c c c c}
\hline
  \hline
 E$_{eff}$       & 0       & 5.18     & 5.24     & 6.17   & 6.79 \\
  \hline
$118.9$  & $ 0.04 \pm 0.03 $ & $                $ & $               $& $0.4  \pm 0.1  $ & $1.2  \pm0.1  $ \\
$124.4$  & $ 0.03 \pm 0.02 $ & $                $ & $               $& $0.43 \pm 0.12 $ & $1.36 \pm0.12 $ \\
$157.0$  & $ 0.015\pm 0.004$ & $ 0.14 \pm  0.01 $ & $0.08\pm 0.01 $& $0.60 \pm 0.03 $ & $1.36 \pm0.05 $ \\
$173.1$  & $ 0.018\pm 0.006$ & $ 0.15 \pm  0.01 $ & $0.10\pm 0.01 $& $0.60 \pm 0.02 $ & $1.26 \pm0.04 $ \\
$182.2$  & $ 0.019\pm 0.004$ & $ 0.147\pm  0.007$ & $0.08\pm 0.01 $& $0.68 \pm 0.02 $ & $1.30 \pm0.03 $ \\
$198.1$  & $ 0.026\pm 0.003$ & $ 0.226\pm  0.009$ & $0.08\pm 0.01 $& $1.00 \pm 0.02 $ & $1.34 \pm0.03 $ \\
$209.5$  & $ 0.032\pm 0.006$ & $ 0.35 \pm  0.01 $ & $0.07\pm 0.01 $& $1.53 \pm 0.03 $ & $1.55 \pm0.03 $ \\
$219.2$  & $ 0.055\pm 0.009$ & $ 0.55 \pm  0.02 $ & $0.12\pm 0.01 $& $2.31 \pm 0.05 $ & $1.94 \pm0.05 $ \\
$229.1$  & $ 0.077\pm 0.005$ & $ 0.95 \pm  0.01 $ & $0.13\pm 0.01 $& $3.57 \pm 0.03 $ & $2.30 \pm0.02 $ \\
$239.1$  & $ 0.177\pm 0.006$ & $ 1.98 \pm  0.03 $ & $0.14\pm 0.01 $& $7.54 \pm 0.07 $ & $3.62 \pm0.05 $ \\
$248.6$  & $ 0.77 \pm 0.02 $ & $ 7.97 \pm  0.06 $ & $0.30\pm 0.03 $& $28.08\pm 0.13 $ & $11.98\pm0.08 $ \\
$256.2$  & $ 7.97 \pm 0.19 $ & $ 93   \pm  2    $ & $               $& $319  \pm 7    $ & $122  \pm3    $ \\
$257.2$  & $ 17.75\pm 0.45 $ & $ 182  \pm  5    $ & $               $& $617  \pm 16   $ & $243  \pm6    $ \\
$258.0$  & $ 43   \pm 1    $ & $ 492  \pm  15   $ & $               $& $1586 \pm 49   $ & $746  \pm23   $ \\
$258.4$  & $ 91   \pm 3    $ & $ 1061 \pm  30   $ & $               $& $3780 \pm 107  $ & $1440 \pm41   $ \\
$258.7$  & $ 169  \pm 5    $ & $ 1719 \pm  55   $ & $               $& $6139 \pm 196  $ & $2532 \pm81   $ \\
$259.0$  & $ 293  \pm 11   $ & $ 2575 \pm  95   $ & $               $& $9488 \pm 351  $ & $4514 \pm167  $ \\
$259.3$  & $ 241  \pm 10   $ & $ 3691 \pm  157  $ & $               $& $10973\pm 467  $ & $4399 \pm187  $ \\
$301.5$  & $ 0.08 \pm 0.01 $ & $ 0.47 \pm  0.06 $ & $               $& $1.49 \pm 0.06 $ & $1.83 \pm0.09 $ \\
$311.4$  & $ 0.073\pm 0.008$ & $ 0.32 \pm  0.03 $ & $               $& $0.99 \pm 0.03 $ & $1.40 \pm0.05 $ \\
$320.9$  & $ 0.063\pm 0.007$ & $ 0.29 \pm  0.02 $ & $               $& $0.64 \pm 0.02 $ & $1.41 \pm0.03 $ \\
$329.1$  & $ 0.060\pm 0.006$ & $ 0.22 \pm  0.02 $ & $               $& $0.39 \pm 0.02 $ & $1.26 \pm0.02 $ \\
$350.3$  & $ 0.045\pm 0.003$ & $ 0.13 \pm  0.01 $ & $               $& $0.24 \pm 0.01 $ & $1.17 \pm0.02 $ \\
$366.8$  & $ 0.043\pm 0.007$ & $ 0.10 \pm  0.04 $ & $               $& $0.23 \pm 0.03 $ & $1.05 \pm0.04 $ \\
 \hline
 \hline
\end{tabular}
\caption{S-factor for different capture transitions. 
         S(E) in units of keV b for transitions to E$_X$ (MeV) states. E$_{eff}$ denotes the effective center of mass energy (in units of keV) within the target \cite{rol88}.
         }
\label{results}
\end{center}
\end{table*}


\begin{table*} 
\begin{center}
\begin{tabular}{ c c  c  c  c c}
\hline
  \hline
  E$_X$ [{\rm keV}] & present& \cite{schroeder87}&\cite{angulo01}&\cite{ru05}&\cite{mukhamedzhanov03}\\
 \hline
  0  & $0.25\pm0.06$     & $1.55\pm0.34$  & $0.08^{+0.13}_{-0.06}$ & $0.49\pm0.08$  & $0.15\pm0.07$  \\
5183 & $0.010\pm0.003$   &                &                        &                &                \\
5241 & $0.070\pm0.003$   &                &                        &                &                \\
6173 & $0.08\pm0.03$     & $0.14\pm0.05$  & $0.06^{+0.01}_{-0.02}$ & $0.04\pm0.01$  & $0.13\pm0.02$  \\
6793 & $1.20\pm0.05$     & $1.41\pm0.02$  & $1.63\pm0.17$          & $1.15\pm0.05$  & $1.4\pm0.2$    \\
\hline                                                                                             
total& $1.61\pm0.08$     & $3.20\pm0.54$  & $1.77\pm0.20$          & $1.68\pm0.09$  & $1.68\pm0.2$   \\
\hline
\end{tabular}
\caption{Values of astrophysical factor S(0) from present and previous work in units of keV b.}
\label{S0}
\end{center}
\end{table*}

\end{document}